# Stabilizing spatially structured populations through Adaptive Limiter Control

**Pratha Sah*[1], Sutirth Dey****

****Corresponding author

**Running Title: Metapopulation stabilization by ALC**

**Affiliations:**

Population Biology Laboratory, Biology Division, Indian Institute of Science Education and Research-Pune, Pashan, Pune, Maharashtra, India, 411 021

1. Present address: Department of Biology, Georgetown University**,** 37th and O Streets, N.W., Washington D.C. 20057, USA

* ps875@georgetown.edu

**Correspondence to Email: s.dey@iiserpune.ac.in

**Name and address of the corresponding author:**

Sutirth Dey

Assistant Professor, Biology Division
Indian Institute of Science Education and Research
3[r]d floor, Central Tower, Sai Trinity Building
Garware Circle, Pashan
Pune - 411 021, Maharashtra, India
Tel: +91-20-25908054

**Abstract**

Stabilizing the dynamics of complex, non-linear systems is a major concern across several scientific disciplines including ecology and conservation biology. Unfortunately, most methods proposed to reduce the fluctuations in chaotic systems are not applicable for real, biological populations. This is because such methods typically require detailed knowledge of system specific parameters and the ability to manipulate them in real time; conditions often not met by most real populations. Moreover, real populations are often noisy and extinction-prone, which can sometimes render such methods ineffective. Here we investigate a control strategy, which works by perturbing the population size, and is robust to reasonable amounts of noise and extinction probability. This strategy, called the Adaptive Limiter Control (ALC), has been previously shown to increase constancy and persistence of laboratory populations and metapopulations of *Drosophila melanogaster*. Here we present a detailed numerical investigation of the effects of ALC on the fluctuations and persistence of metapopulations. We show that at high migration rates, application of ALC does not require *a priori* information about the population growth rates. We also show that ALC can stabilize metapopulations even when applied to as low as one-tenth of the total number of subpopulations. Moreover, ALC is effective even when the subpopulations have high extinction rates: conditions under which one other control algorithm has previously failed to attain stability. Importantly, ALC not only reduces the fluctuation in metapopulation sizes, but also the global extinction probability. Finally, the method is robust to moderate levels of noise in the dynamics and the carrying capacity of the environment. These results, coupled with our earlier empirical findings, establish ALC to be a strong candidate for stabilizing real biological metapopulations.

**Author summary**

Spatially-structured populations (metapopulations) that fluctuate widely in size or have a high probability of extinction, pose major challenges for conservation biologists. Most methods proposed to stabilize the dynamics of such unstable metapopulations have been shown to reduce either the fluctuations in sizes or the extinction probability, but seldom both. Recently, we proposed and empirically verified a stabilization method (called the Adaptive Limiter Control or ALC) that can not only decrease the temporal variations in size, but also enhance the persistence of populations and metapopulations. However, the question remained as to whether the strategy would still be effective under situations not investigated in the previous experiments. Here we study the efficiency of ALC in stabilizing metapopulations under a wide range of conditions through biologically realistic simulations. We show that ALC is robust to factors such as high local extinction probabilities, and different environmental carrying capacities or population growth rates. The method works even when as low as one-tenth of the constituent populations are perturbed. Since we use a widely –applicable model of population dynamics, our results are likely to be generalizable over large number of species. Overall, we establish ALC as a strong candidate for stabilizing real, biological metapopulations.

# Introduction

Controlling erratically fluctuating and extinction prone populations is of major interest to ecologists and conservation biologists, and has been an active area of investigation for the last two decades [1]. Although substantial progress has been made in terms of control theory in the fields of chemical sciences, physical sciences, electrical engineering, medicine and economics (reviewed in [2,3]), much of the insights gained from such studies are not applicable to the problem of controlling biological populations. This is partly because majority of these methods pertain to the amelioration of chaos to obtain stable points or trajectories of specific periodicities. However, short and noisy time series typical of most biological populations make it statistically difficult to distinguish noisy limit cycles from chaotic trajectories. Furthermore, although many chaos control strategies involve perturbing the parameters of the system, the parameters of biological populations (e.g. intrinsic growth rate or carrying capacity) are typically estimated *a posteriori* through model-fitting, and are almost never available for perturbation. Thus, most strategies proposed in the control-theory literature cannot be directly applied to stabilize the dynamics of biological populations. To alleviate some of the above-mentioned problems, a number of methods have been proposed that induce stability through perturbation of the state-variable, i.e. the population size [4,5,6,7,8]. For example, it has been theoretically shown that constant immigration can convert chaotic trajectories into limit cycles in spatially homogeneous [7,9] as well as spatially-structured [4,6] populations. Similarly, regular perturbation towards a target population size can also reduce the overall temporal fluctuations in the time series [10].

Another promising chaos control strategy in the context of biological populations is the so-called "limiter*"* family of algorithms. Broadly speaking, the limiter strategy works by not allowing the population size to go above or below some pre-determined threshold, and typically requires some *a priori* information about the dynamics of the system. Although proposed and verified in the context of physical systems like chaotic mechanical pendulums and double diode circuits [11], later theoretical investigations have established the generalizability of the concept to other systems (e.g. [12]) including models of population dynamics [13,14]. However, until recently, there was no empirical support for the efficacy of any of the several limiter control algorithms in the context of biological populations or metapopulations.

Recently, we proposed a novel limiter strategy, called the Adaptive Limiter Control (ALC), to stabilize the dynamics of spatially-unstructured and –structured populations [15]. ALC is a restocking strategy that seeks to maintain populations and metapopulations above a threshold. However, instead of a fixed threshold [13,14], the magnitude of the perturbation is a function of the difference in the population size in two successive generations. This ensures that little *a priori* knowledge is needed in terms of the dynamics of the system to implement this method. We also empirically demonstrated the effectiveness of ALC in reducing the amplitude of fluctuations in size of replicate laboratory populations and metapopulations of *Drosophila melanogaster*. Interestingly, ALC was also able to reduce the extinction probability of the said populations as well. Biologically realistic simulations of ALC using three different non-species-specific models that also incorporated noise, extinction probabilities and lattice effects [16], were able to capture most of the trends of the data. These showed that the obtained results were not due to some idiosyncratic features of the experimental system but are likely to be generalizable. However, owing to its emphasis on empirical validation, the previous study [15] lacked a systematic exploration of ALC, particularly in terms of effects on the dynamics of spatially-structured populations. We aim to fulfil that lacuna here.

In this study we further explore the efficacy of ALC in stabilizing the chaotic dynamics of metapopulations governed by coupled Ricker maps. We demonstrate that Adaptive limiter control (ALC), successfully achieves global stability over a wide range of parameters as compared to an unperturbed system. The controller is shown to be robust to various biological realities such as varying subpopulation numbers, carrying capacity, intrinsic population growth rate and extinction probability in a metapopulation. We find that, barring low migration rates, ALC either stabilizes metapopulations or fails to have an effect, but never reduces the global stability as compared to its uncontrolled counterpart. This implies that one does not require extensive *a priori* knowledge of the parameter values of the metapopulations, which is a definite benefit in terms of the applicability of the method for practical purposes.

## Results and discussion

**Effects of migration rate on constancy**

The rate of migration between subpopulations is known to influence the dynamics of metapopulations [17,18,19] and thereby can have a major impact on the efficacy of a control technique [15]. We therefore tested the effect of ALC on the constancy of two-patch metapopulations at different rates of symmetric migration. We find that compared to the controls ($c = 0$), both LALC ($c = 0.25$) and HALC ($c = 0.4$), seem to increase the Fluctuation Index (FI) of the metapopulation at low (<20%) rates of migration (Fig.1A). However, when migration rates are higher, the situation reverses and both levels of ALC seem to reduce the metapopulation fluctuation index. The explanation for this phenomenon perhaps lies in the way ALC affects the synchrony between the constituent subpopulations (Fig 1B). In an unperturbed system ($c = 0$), low rates of migration (< 20%) reduces metapopulation FI by inducing out-of-phase fluctuations (i.e. negative synchrony) between neighboring sub-populations [17]. This happens because negative synchrony ensures that crashes in some subpopulations are accompanied by booms in others. This in turn reduces the temporal variation in the metapopulation size [17,18,19], which by definition, is the sum of all subpopulation sizes. Conversely, in-phase fluctuations between subpopulations at high migration rates reduce constancy (i.e. increase FI) at the metapopulation level, by bringing the subpopulations in phase with each other. It has been earlier shown that ALC reduces both positive and negative synchrony between subpopulations [15], which is expected to have contrasting effects on constancy stability. While the reduction of positive synchrony (Fig 1B) reduces metapopulation FI, the lowering of negative synchrony at low migration rate actually increases it, leading to the observed opposite effects of ALC at the two migration rates (Fig 1A).

It is clear from Fig 1A that an unperturbed system has high FI at higher rates of migration. Since ALC is a perturbation strategy to stabilize an unstable system, we focus on a particular high rate of migration (=30%) for the rest of our investigation.

**Effects of *r* on constancy and persistence**

Estimating the precise values of parameters like growth rate or carrying capacity is typically difficult for any real population, and often a control strategy will need to be applied without much prior information about the dynamics of a system. Under such conditions, a control method that is known to stabilize metapopulations only under a narrow parameter zone, is expected to be of limited use. As part of our investigations on its applicability, we tested the efficacy of ALC in terms of both constancy and persistence at various magnitudes of *r* for the subpopulations (Fig 2). HALC reduces metapopulation FI for all values of *r* > 2.2 where as LALC is effective at a slightly higher range of *r* (> 2.7) (Fig. 2A). ALC had no discernible effect on the dynamics at *r* < 2.2. These observations can perhaps be explained by an interaction of the nature of ALC and the Ricker dynamics. The Ricker model is known to follow a period-doubling route to chaos with the amplitude of oscillation of the population size becoming larger with increasing *r* [20]. ALC perturbations happen only when the population size in a given generation is less than a fraction *c* of its previous generation. In the Ricker model, *r* < 2.0 always leads to a stable point cycle, as a result of which, the ALC perturbation is never applied, and the dynamics of the unperturbed population is indistinguishable from the ALC-controlled ones. When *r* lies between 2.0 and 2.2, the system undergoes small amplitude two-point limit cycles in which the population crashes are not sufficiently large to lead to the application of ALC. Thus, again, there is no difference between the control and the perturbed populations. It is only when the amplitude of the limit cycles become sufficiently large (*r* > 2.2 in this case) that ALC perturbations are actually applied and there is an effect of the perturbation on FI. Not surprisingly, this reduction of FI is visible for lower values of *r* for HALC (*r* > 2.2) compared to LALC (*r* > 2.7). This is because the minimum amplitude of the crash needed for ALC to be applied is 60% and 75% of the previous population size for HALC and LALC respectively. This implies that compared to LALC, HALC perturbations starts happening at lower values of *r* and therefore the stabilizing effect also manifests earlier (Fig 2A). Evidently, these observations should be applicable for all models that follow a period-doubling route to chaos like the logistic [21] or the Hassell [22], although the exact values of the growth rate parameter where ALC becomes effective as a stabilizing factor, would differ. Since the nature of the dynamics (i.e. stable point, limit cycle or chaos) of a Ricker model is independent of the carrying capacity (*K*), our reasoning suggests that the effect of ALC should also be unaffected by the values of *K*, which was indeed found to be the case (Fig. S1). The interaction between the intrinsic growth rate and the ALC magnitude in determining the metapopulation FI is investigated more thoroughly in Fig S2.

We also investigated the persistence of ALC-perturbed metapopulations, as preventing extinction can be a more pressing objective under certain scenarios. It is tempting to reason that constancy and persistence are correlated because, all else being equal, populations that fluctuate more are expected to hit lower values more often over a given length of time. Each time a populations hits one of these low sizes, it becomes more vulnerable to demographic stochasticity, thus increasing the probability of extinction. Although intuitive, this line of reasoning fails under a number of scenarios. For example, if the population crashes to values that are still somewhat high, then the effect of demographic stochasticity would be comparatively less, leading to a breakdown in the correlation between constancy and persistence [23]. Therefore, we explicitly investigated the ability of ALC to induce metapopulation persistence. Although ALC was in general effective in reducing the global extinction probability, we did not find a good correspondence between FI (Fig 2A) and the corresponding extinction probability (Fig 2B) for different values of $r$. This was consistent with previous empirical findings that indicate constancy and persistence are often uncorrelated [24,25]. But then, if not constancy, what leads to this enhanced persistence of ALC-controlled populations? The answer perhaps lies in the ability of ALC to reduce positive synchrony [15], which reduces the extinction probability of connected subpopulations [26,27]. This is because a high positive synchrony between populations causes the populations to reach lower levels simultaneously, reducing the chances of each population to receive immigrants from neighboring populations and thereby increasing the chances of local (and global) extinction. A reduction in positive synchrony by ALC desynchronizes the fluctuation of neighboring populations, which ensures that whenever a population reaches a low population size, it is often "rescued" from extinction by receives immigrants from neighboring population with higher population size.

**Local extinction probability and constancy**

The subpopulations of a metapopulation often go extinct [28,29], which in turn can play a role in determining the dynamics of the system [30,31]. Such local extinctions can also modulate the effects of a control strategy. For example, local extinctions were implicated when constant immigration [6,32] were found ineffective in stabilizing laboratory metapopulations of *Drosophila melanogaster* [33]. Although ALC has been empirically demonstrated to be effective in stabilizing the dynamics of extinction-prone populations [15], a detailed investigation of the effects of local extinction on metapopulation stability is lacking. In the simulations of the previous section, we had assumed a particular probability of extinction (= 0.5) when a subpopulation fell below the critical population size threshold of 4. We therefore investigated the performance of ALC for different values of the extinction probability of the subpopulations each time they went below a threshold of 4. We found that increasing the probability of subpopulation extinction did not reduce the ability of ALC to induce metapopulation stability (Fig. 3A). This observation also held true on varying the threshold of critical population sizes keeping a constant extinction probability of 0.5 (Fig 3B). The robustness of ALC towards subpopulation extinction probability can be explained by the very way in which ALC is designed: in the time series whenever there are population crashes (including extinctions),

ALC brings the population size back to a higher number. Thus extremely low values of the breeding-population size are never permitted irrespective of the extinction probability or the critical threshold. This reduces the magnitude of both the population crashes as well as the subsequent spikes, which in turn contributes to the reduction in Fluctuation Index.

**Larger metapopulations and larger fraction of controlled patches**

So far we have investigated the adaptive limiter control mechanism on a simple metapopulation consisting of only two subpopulations. However, it is known that metapopulation dynamics can be considerably influenced by the number of constituent subpopulations [34]. Moreover, in our study, we have perturbed 50% of the subpopulations (*i.e.* one out of two), a fraction which might be difficult to achieve in larger metapopulations in practise. We therefore tested the ability of ALC to stabilize larger metapopulations with only one of the constituent subpopulations being subjected to ALC control. Given that only one patch in the whole metapopulation is being controlled, intuitively, the efficacy of ALC should go down sharply as the total number of subpopulations increases. Surprisingly, ALC perturbed metapopulations with up to 10 subpopulations were still more stable than their unperturbed counterparts (Fig 4A and 4B), with the effect being most pronounced for number of patches $\leq 5$. The FI of an ALC-controlled metapopulation becomes equivalent to an unperturbed metapopulation when there are more than ten subpopulations. This is perhaps because at that level, the fraction of perturbed subpopulations ($< 0.1$) becomes too less to affect the dynamics at a global scale. Perturbing large number of subpopulations with LALC should be avoided as the global FI increases with increasing number of patches for $c \sim 0.3$ (Fig 5). This observation is consistent with an earlier finding that under constant immigration, increasing the fraction of perturbed patches leads to an increase in metapopulation FI [33]. This implies that the efficacy of a control algorithm can be context-specific and "more control" does not always translate into "better control".

**Closing Remarks**

The dynamics of spatially-connected populations are crucially dependent on the life-history of the organisms [35], which in turn can potentially affect the efficacy of a control algorithm. In this paper, we explore this aspect and show how the magnitude of the control parameter (*c*) interacts with the intrinsic growth rate, to determine the constancy and persistence stability of the metapopulation. At a practical level, our main message is that as long as migration rates are high, ALC can either enhance the metapopulation constancy and persistence or have no effect, but can never reduce the stability of the metapopulation to a level less than that of the corresponding control with unstable dynamics. This indicates that precise *a priori* knowledge of the growth rate / localized extinction rates and carrying capacity of the constituent subpopulations are not really needed for the application of ALC. The generalizability of our results is also enhanced by the fact that we used the Ricker model which has been shown to be a good descriptor of the dynamics of populations including bacteria [36], fungi [37], ciliates [38], crustaceans [39], fruit flies [40,41], fishes [42] etc. This remarkable success of the Ricker model is probably because derivations from first-principles show that populations that exhibit scramble competition and random distribution of the individuals are expected to follow the Ricker dynamics [43]. However, several organisms that appear on conservation lists (like reptiles, birds, mammals) may not follow these assumptions or their dynamics may not be well-represented by the Ricker model. Therefore, ALC should not be applied to such species, until and unless it is shown to have a stabilizing effect in the context of the appropriate dynamics. Moreover, the scheme of migration [44] and the form of density dependence [45] are known to affect metapopulation dynamics, two factors which were not investigated in this study. Therefore, any attempts to use ALC to control real populations should be based on relevant information about the biology of the organism.

## Methods

**Adaptive Limiter Control (ALC)**

ALC stabilizes populations by preventing a (sub) population from going below a predefined fraction ($c$) of its size in the previous generation. Since $c$ is a fraction and not a hard number, the method automatically "adapts" to populations inhabiting environments with different carrying capacities or exhibiting a regular upward or downward trend. The population is perturbed only if the current population size falls below the ALC threshold and involves restocking individuals from an external source till the current population size reaches the ALC threshold. The model can be thus represented as:

$$N^*_t = N_t \qquad \text{if } N_t \geq c.N^*_{t-1} \qquad (1)$$

$$N^*_t = c.N^*_{t-1} \qquad \text{if } N_t < c.N^*_{t-1} \qquad (2)$$

where $N$ indicates the population size at a particular generation before the imposition of ALC, $N^*$ is the population size post ALC treatment and $c$ is the ALC magnitude. Note that $N^*$ is also the breeding population size at the end of a generation. Therefore, the population size of the $t+1$ th generation before ALC treatment will be $N_{t+1} = f(N^*_t)$. It is known that the time series of $N_t$ and $N^*_t$ can potentially have different dynamics [46]. Here we restrict ourselves to the dynamics of $N_t^*$ since being the breeding population size, it is more relevant from a biological point of view.

Evidently when $c = 1$ and ALC is implemented in every constituent population of a metapopulation, it has the potential of reducing the dynamics of the system to a fixed point. However, this is practically difficult to achieve due to the massive intervention effort that would be needed. Thus, we focus on the stabilizing efficacy of much lower values of $c$ applied to only a subset of subpopulations. Following a previous study [15], the rest of our analysis and discussion focuses on two values of ALC: $c = 0.25$ and $c =$ 0.4 which we refer to as Low Adaptive Limiter Control (LALC) and High Adaptive Limiter Control respectively.

**Simulations**

We used the Ricker equation [42] to examine the asymptotic behaviour of ALC and the most effective value of *c*. Ricker equation is given as ( $N_{t+1} = N_t . e^{r.(1-\frac{N_t}{K})}$ ) where $N_t$ denotes the population size at time *t*, *r* is the per-capita intrinsic growth rate and *K* is the carrying capacity. On each iteration noise was added to the *r* parameter in the form of a random number ε drawn from a uniform random distribution of range -0.2≤ε≤ 0.2. The final population growth model can thus be represented as: $N_{t+1} = N_t.\exp((r+\varepsilon).(1-N_t/K))$. In our simulations we consider the value of intrinsic growth rate parameter to be 3.5 which lies within the range of parameter space representing chaotic region in a Ricker map. The initial population size ($N_0$) and the carrying capacity (*K*) were fixed at 20 and 30 respectively.

The unmodified Ricker model does not takes zero-values and thus theoretical populations (and metapopulations) governed by the Ricker population growth function never go extinct. As subpopulation extinction is known have an impact on the dynamics of metapopulations [17,33], we explicitly introduced stochastic extinction in our models by implementing an extinction probability of 0.5 below a threshold population size of 4 [15,17]. On the event of metapopulation extinction, the metapopulation was reset with a population size of 8 per subpopulation. Note that for *c* > 0, implementation of ALC ensures that the metapopulation size is more than zero at each generation. Thus, only the unperturbed (*c* = 0) metapopulation were reset during extinction events.

For simulations in this study, a metapopulation is described as two or more subpopulations connected to each other via symmetric rate of migration. For metapopulations consisting of more than two subpopulations, the subpopulations were considered to occupy spaces on the periphery of an imaginary circle so that dispersal occurred between the two nearest neighbours of a subpopulation (i.e. linearly arranged with periodic boundary condition). In nature, such metapopulations can be found along the edge of a lake or a park. In perturbed metapopulations, ALC was imposed after migration so that immigration due to ALC for a given generation would have an impact on the population size of the neighbour only through migration in the subsequent generation.

All simulations were run using MATLAB ® R2010a (Mathworks Inc.) and each point in the figures represents an average of 100 simulation runs. The error bars represent stand error of the mean. The first 400 iterations of each run were rejected, and all indices were computed over the next 100 iterations.

**Measure of constancy, persistence and synchrony**

We considered two attributes of stability in this study: constancy and persistence. Constancy is defined as the property of a system to stay essentially unchanged [47]. We quantified constancy using a widely used [48,49] measure called the Fluctuation Index (FI);

$$FI = \left[\sum_{t=0}^{t-1} abs(N_{t+1} - N_t)\right] / (\bar{N} \times T)$$

FI is a dimensionless quantity which measures one-step fluctuation in population or metapopulation size across generations, scaled over mean population size [17]. High FI implies reduced constancy and vice versa.

Persistence was quantified as metapopulation extinction frequency i.e. the average number of generations a metapopulation records a zero population size (before the application of ALC) as a proportion of the total number of generations. Thus, high extinction frequency of the system indicated the system to be less persistent. We calculated synchrony as the cross-correlation coefficient at lag zero of the first-differenced log-transformed values of the two subpopulation sizes [50].

## Acknowledgements

The authors thank Sudipta Tung for helpful discussions and comments on the manuscript.

**FIGURES**

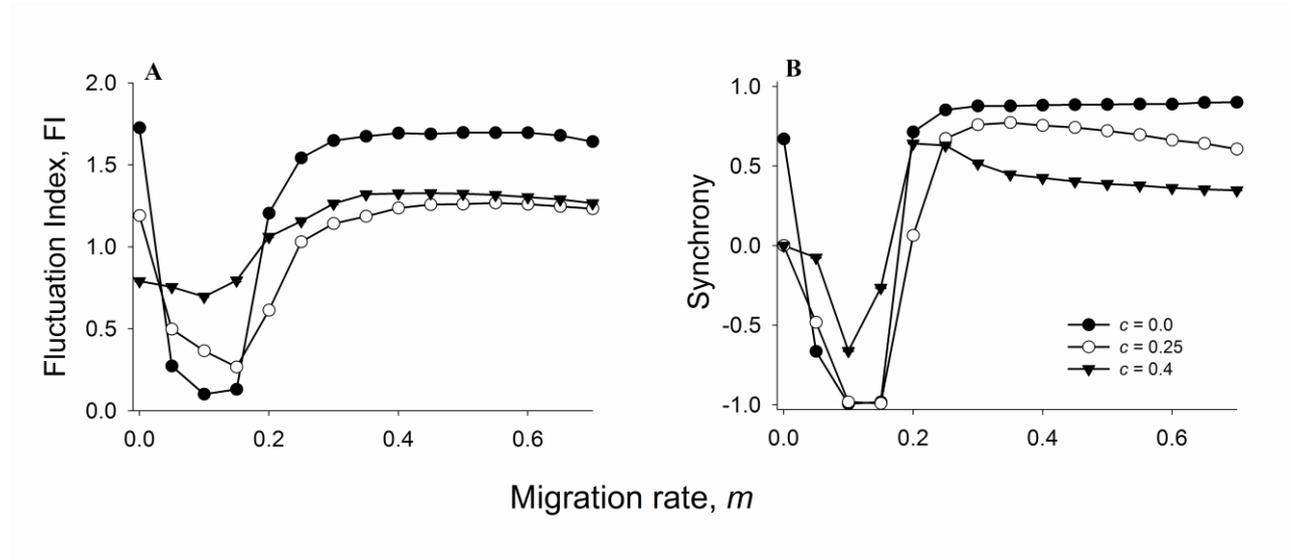

Figure 1: **Effects of ALC on metapopulation FI and synchrony at different rate of migration.** (**A**). Both LALC ($c = 0.25$) and HALC ($c = 0.4$) increases metapopulation FI at low migration rates, but reduces the same at high migration rates. This contrasting effect can be explained by (**B**) which shows that ALC reduces both positive and negative synchrony, which in turn is expected to have opposite effects on metapopulation constancy. Each point in this and all subsequent figures is a mean of 100 independent runs. Error bars denote ±SEM.

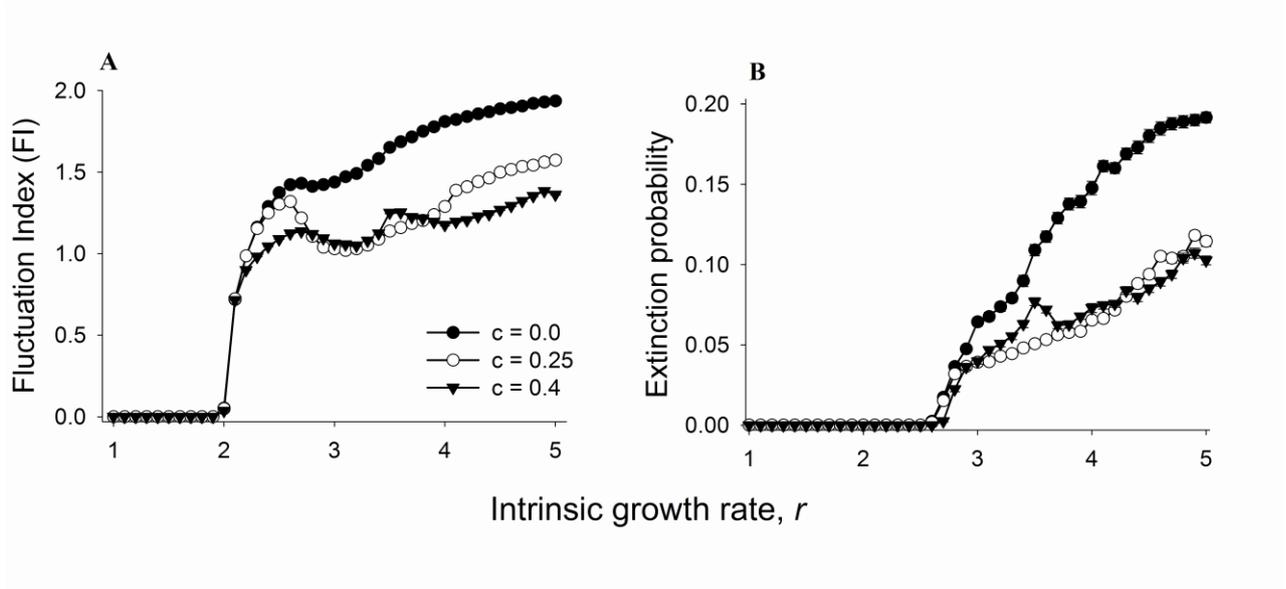

Figure 2: **Effects of ALC on metapopulation stability at different intrinsic growth rate (*r*) values**. ALC enhances (**A**) constancy and (**B**) persistence over a wide parameter range, and has no effects in other zones. See main text for a possible explanation.

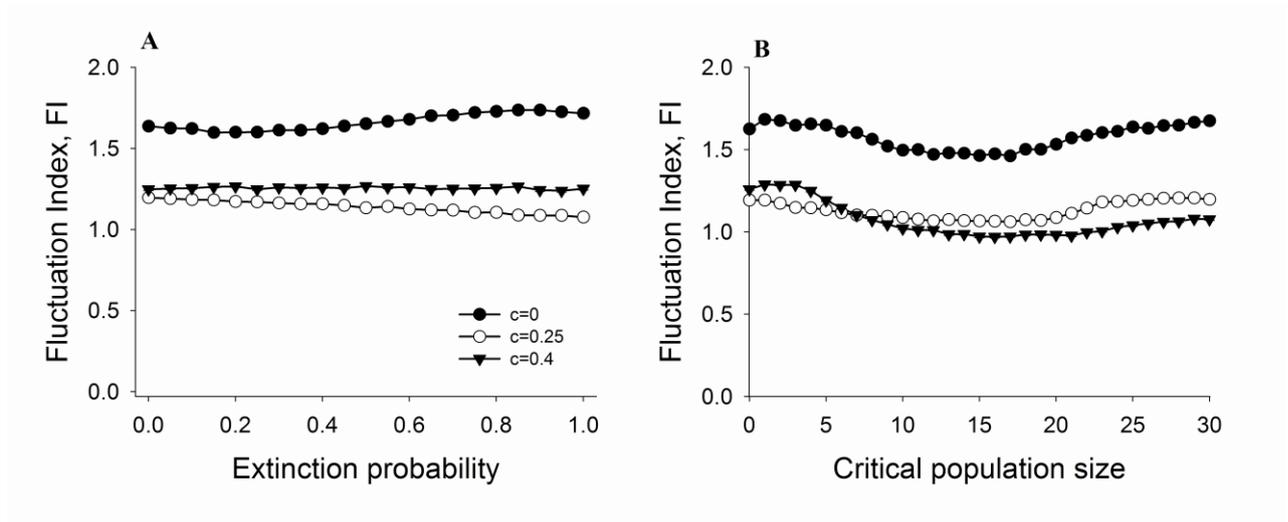

Figure 3: **Effects of ALC on metapopulation constancy under different rates of subpopulation extinction.** (**A**). With increasing extinction probability when the population size goes below 4. (**B**) With increasing critical population sizes below which, there was a 50% extinction probability that the population would go extinct. In both cases, increasing the rate of extinction did not reduce the efficacy of ALC in inducing greater constancy. See main text for a possible explanation.

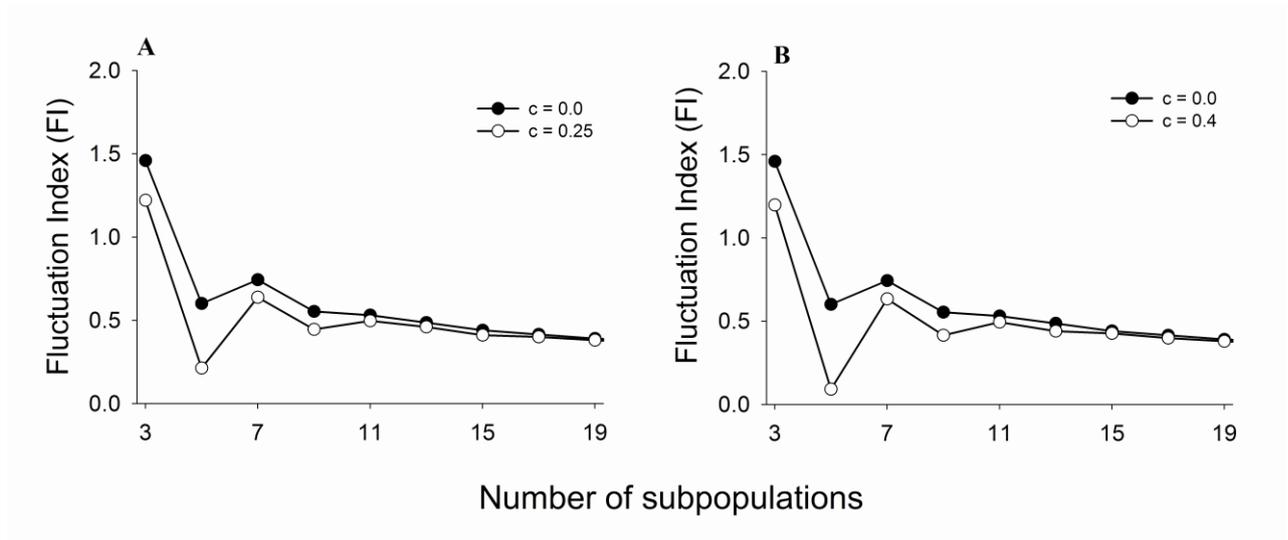

Figure 4: **Effects of ALC on constancy in metapopulations with different number of subpopulations.** (**A**) LALC (i.e. c = 0.25), and (**B**) HALC (i.e. c = 0.4). In both figures, only one subpopulation is perturbed for increasing number of subpopulations. Perturbing only 1 patch by ALC can reduce FI of metapopulations with up to 10 subpopulations.

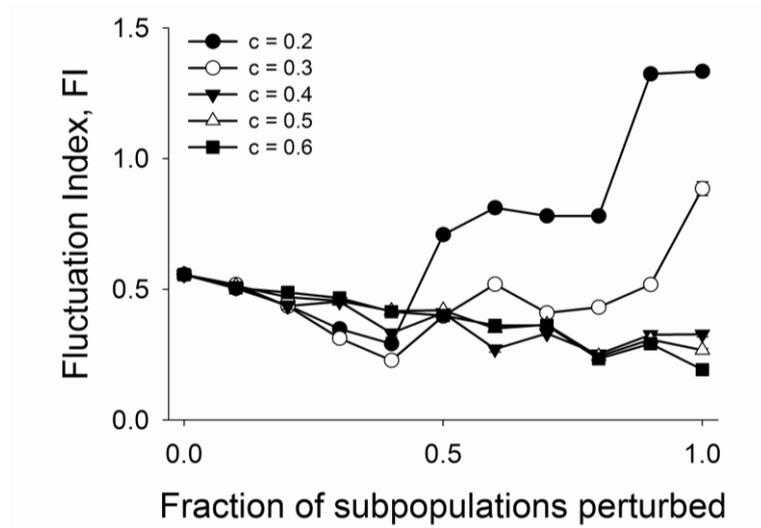

Figure 5: **Effects of increasing the fraction of ALC controlled subpopulation on metapopulation constancy.** In this figure, each metapopulation consists of 10 subpopulations. For low values of $c$, increasing the fraction of perturbed subpopulations can have a negative effect on constancy.

**SUPPLEMANTARY ONLINE MATERIAL**

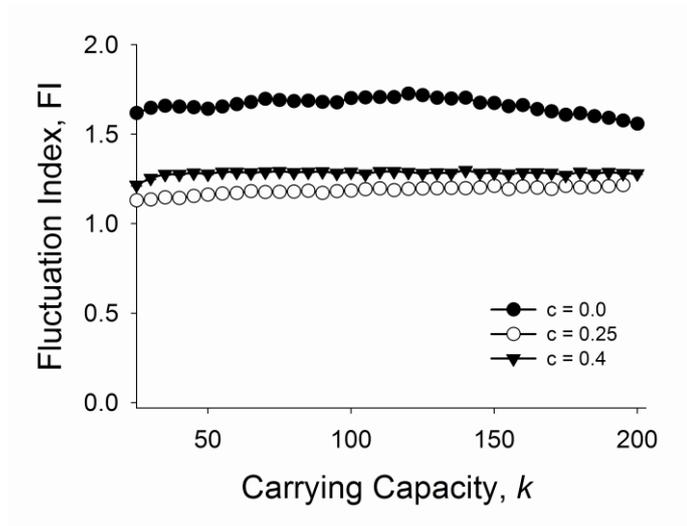

Figure S1: **Effects of ALC on metapopulation constancy at different magnitudes of carrying capacity.**

There was no effect of carrying capacity on the stabilizing efficiency of ALC.

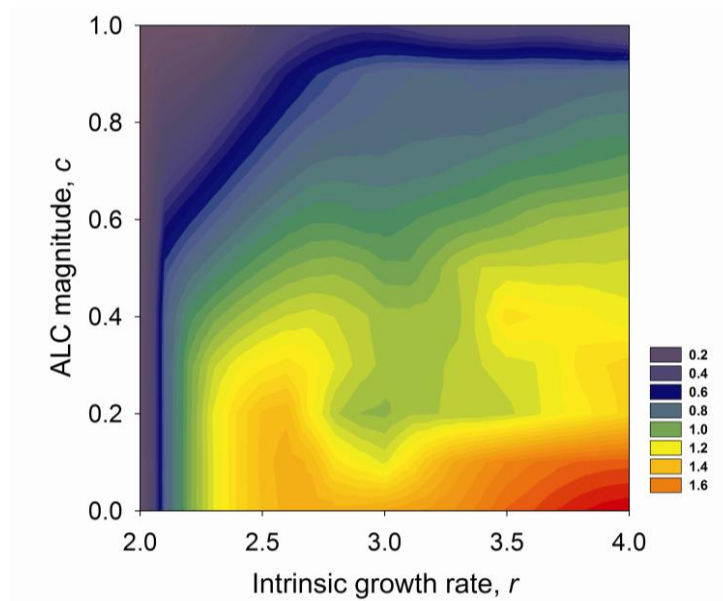

Figure S2: **FI of metapopulation with 2 subpopulation as a function of intrinsic growth rate ($r$) and ALC magnitude ($c$).**